# Equation-Free Computations as DDDAS Protocols for Bifurcation Studies: A Granular Chain Example

M. O. Williams[1], Y. M. Psarellis[3], D. Pozharskiy [2], C. Chong [4], F. Li [5], J. Yang [6], P.G. Kevrekidis[4], I.G. Kevrekidis [3]

[1] Oceanit Laboratories, Honolulu, Hawaii 96813, USA
[2] Department of Chemical and Biological Engineering and PACM, Princeton University, Princeton, New Jersey 08544, USA
[3] Department Of Chemical and Biomolecular Engineering, Johns Hopkins University, MD, USA
[4] Department of Mathematics and Statistics, University of Massachusetts, Amherst, Massachusetts 01003-4515, USA
[5] Graduate Aerospace Laboratories (GALCIT), California Institute of Technology, Pasadena, California 91125, USA
[6] Aeronautics and Astronautics, University of Washington, Seattle, Washington 98195-2400, USA

**Abstract.** This chapter discusses the development and implementation of algorithms based on Equation--Free/Dynamic Data Driven Applications Systems (EF/DDDAS) protocols for the computer-assisted study of the bifurcation structure of complex dynamical systems, such as those that arise in biology (neuronal networks, cell populations), multiscale systems in physics, chemistry and engineering, and system modeling in the social sciences. An illustrative example demonstrates the experimental realization of a chain of granular particles (a so-called engineered granular chain). In particular, the focus is on the detection/stability analysis of time-periodic, spatially localized structures referred to as "dark breathers". Results in this chapter highlight, both experimentally and numerically, that the number of breathers can be controlled by varying the frequency as well as the amplitude of an "out of phase" actuation, and that a "snaking" structure in the bifurcation diagram (computed through standard, model-based numerical methods for dynamical systems) is also recovered through the EF/DDDAS methods operating on a *black-box* simulator. The EF/DDDAS protocols presented here are, therefore, a step towards general purpose protocols for performing detailed bifurcation analyses directly on laboratory experiments, not only on their mathematical models, but also on measured data.

# 1 Introduction

## 1.1 Overview

This chapter discusses the development and implementation of Dynamic Data Driven Applications Systems (DDDAS) protocols for the Equation-Free (EF) computer-assisted study of the bifurcation structure of complex dynamical systems. Bifurcation studies are an invaluable tool for understanding how a nonlinear system's dynamics change over a range of parameters including: (i) how a given solution changes as the system parameters change, (ii) when these solutions are robust to perturbations and when they are not, and (iii) what new types of behaviors are expected to emerge from an unstable solution. As a result, bifurcation analysis is commonly performed in any setting that can be modeled or approximated as a system of coupled ODEs.

In this chapter, two bifurcation studies are performed on an *Engineered Granular Chain (EGC)*, which consist of closely packed arrays of particles [1, 2] that interact elastically. The first study follows the standard approach where a set of governing equations in the form of coupled ODEs are derived, their predictions are validated against experimental data, and the system of equations are used to compute a bifurcation diagram using standard tools like AUTO [3], MATCONT [4], or LOCA [5]. The second bifurcation study uses an EF/DDDAS approach. In this approach, it is assumed that explicit governing equations are not available. Instead, a modified set of *equation-free* algorithms are used that leverage DDDAS ideas to obtain needed information by querying the underlying experimental system or black-box simulator. The objective of this dual approach is to demonstrate that EF/DDDAS bifurcation techniques can obtain the same information generated by more typical equation-aware methods on a realistic and complex system.

## 1.2 Dark-Breathers in Engineered Granular Chains

Engineered Granular Chains (EGCs) were chosen as a test system of EF/DDDAS techniques both for their relevance in applications and because they produce interesting dynamics such as *bright-* and *dark-breathers.* EGCs have been used in applications such as shock and energy absorbing layers [6, 7, 8, 9], acoustic lenses [10], acoustic diodes [11] and sound scramblers [12, 13]. Furthermore, breather solutions in particular have been proposed as a novel energy harvesting device [11].

A soliton is a solitary wave that behaves like a "particle", in that it satisfies the following conditions: (i) It must maintain its shape when it moves at constant speed and (ii) when a soliton interacts with another soliton, it emerges from the "collision" unchanged except possibly for a phase shift [14]. A breather is a localized wave that is periodic in time or space [15, 16, 17, 18]. Bright-breathers are time periodic solutions with high amplitudes at localized regions in space that decay to zero outside of these regions (e.g., $\alpha \operatorname{sech}(x) \cos(2\pi f t)$), and are named such because in optics applications they would appear as a bright spot on a darker background. On the other hand, dark-breathers are the opposite and are time periodic solutions with localized regions of low amplitude (e.g., $\alpha \tanh(x) \cos(2\pi f t)$), and would appear as a dark spot on a brighter background.

Dark-breathers have received far less attention than their bright counterparts, but have been observed in many applications. These include: surface water waves [19], Bose-Einstein condensates [20, 21], ferromagnetic film strips [22] or optical waveguide arrays [23]. They have also attracted attention in quantum mechanical applications including: metamaterials [24], optical fibers [25], graphene lattices [26] and DNA conformations [27].

## 1.3 Equation Free Bifurcation Studies

As mentioned previously, bifurcation studies often seek understand how solutions change as system parameters change, the robustness of these solutions to perturbations, and what other behaviors may arise when a solution becomes unstable. These questions cannot be answered analytically in all but the simplest of systems, so instead a model of the system in the form of a coupled ODE is studied numerically.

Given such a model, packages like AUTO determine how solutions change as system parameters change using *numerical continuation* techniques like pseudo-arclength continuation. Pseudo-arclength continuation computes *branches of solutions* by iteratively finding solutions that are a given distance along the branch of solutions (i.e., an approximation of arclength) away from the initial point [3]. The robustness of a solution to small perturbations can be determined through *linear stability analysis*, which creates a linear model of the system around some target point and studies the dynamics of that linear model to understand the system's response to perturbations. Finally, if a solution is unstable *computing the unstable manifold* of the solution can help to determine what other behaviors will be observed in lieu of the unstable one.

To accomplish these goals, standard tools require access to the explicit governing equations in the form of coupled ODEs so that associated information such as temporal information and derivatives with respect to states and/or parameters can be computed. The models themselves are obtained through a variety of sources including derivation from physics, completely empirically models obtained by fitting a machine learning model to the data, and approaches with traits of both, such as reduced order models or parameter estimation using ensemble methods in data assimilation [28].While having a model enables the use of mature tools such as AUTO, the usefulness of the resulting analysis will be determined by the accuracy of the model as the study is performed on the model and not the underlying system.

The objective of Equation-Free (EF) analyses is to *perform system-level without an explicit model* [29, 30, 31, 32], and are intended for applications where explicit governing equations are not available. This includes multi-scale applications where equations are available on fast timescales but not on slow timescales to enable long-time behaviors to be studied, cases where the underlying system is an agent-based model and not a system of ODEs, and cases where the inner workings of the best available simulator are not available for examination. EF techniques have been used in multiple applications including biology (neuronal networks, cell populations) [33, 34], multiscale systems in physics, chemistry and engineering [30, 35, 36, 37] and even system models in the social sciences [38, 39].

EF/DDDAS techniques in particular obtain the information that would be found by running or differentiating the model by designing experiments that can be run either physically or on a black-box simulator. To minimize the number of experiments that need to be run, EF/DDDAS methods must use different algorithms than found in packages in AUTO. In particular, these methods favor techniques from *matrix-free linear algebra* which use sequences of matrix-vector products in lieu of explicit matrix decompositions [40, 41].

The remainder of the chapter is structured as follows: Section 2 described the EF/DDDAS approach and the experimental setup, Section 3 presents the results obtain from applying a standard, model-based approach, and Section 4 shows an equivalent set of results obtained without an explicit equation or model using EF/DDDAS methods. Finally, Section 5 contains brief conclusions and possible avenues for future research and development.

## 2 Theoretical Method and the Experimental Setup

### 2.1 The EF/DDDAS approach

When the system evolution equations are available, the most popular software choices for bifurcation analysis are AUTO [3], which could perhaps be considered as the industry standard, MATCONT [4], which interfaces with MATLAB, and (intended for large scale computations) the Library of Continuation Algorithms [5], which is a part of the Trilinos project. Apart from numerical continuation these software packages can perform linearized stability analysis and can compute the low-dimensional unstable (sub)manifolds of a given unstable (saddle-type) solution.

The ultimate goal of any bifurcation study is to uncover and quantitatively summarize the dynamics of real world systems. There are many types of useful analyses, but in this section the focus is on EF/DDDAS protocols to solve three common tasks in bifurcation analysis:
1. Numerically computing a solution branch;
2. Computing and determining the stability of the solutions on the branch; and
3. Approximating the unstable (sub)manifolds of given solutions.

To accomplish these tasks, EF/DDDAS algorithms assume:
1. The underlying system has the form: $x(i+1) = F(x(i), p)$ where $x$ is the system state and $p$ the parameters.
2. The system is a black box so neither $F$ nor its derivatives have explicit equations;
3. Initial conditions can be specified at will and the system state measured at any subsequent point in time.

The motivation for these assumptions is to mimic the constraints of both black box simulators and well-controlled experiments where the system dynamics can be observed but not directly manipulated. However, the limitations they impose conflict with the algorithms for bifurcation analysis implemented in packages like AUTO, which require access to derivative information such as Jacobians and Hessians.

To circumvent these limitations, EF/DDDAS protocols leverage matrix-free methods in linear algebra. *Matrix Free* methods were originally designed to solve the linear system $Ax = b$ using only matrix–vector products (i.e., $Ax, Ax^2$ ...) rather than decomposing the full matrix [40, 41]. For example, suppose the objective is to solve the algebraic system of equations $f(x) = 0$. If $x(i)$ is the $i$-th iterate of the Newton method, then $x(i+1) = x(i) + \delta x$ where $J \cdot \delta x = -f(x(i))$. A matrix-free approach to compute $\delta x$ is the Generalized Minimal Residual Method (GMRES) [42], which only requires Jacobian–vector products. Because EF/DDDAS methods lack access to derivative information, these products are approximated as:

$$J \cdot v = \frac{f(x + \varepsilon v) - f(x)}{\varepsilon} + O(\varepsilon), \qquad (1)$$

for $\varepsilon \ll 1$ and $\|v\| = 1$. If the underlying system, $f$, is computational in nature, GMRES can locate equilibrium points with high accuracy, but when the underlying system is a physical experiment the resulting solution will be an approximation whose accuracy depends on how accurately $f$ can be estimated.

### 2.1.1 Numerically Computing Solution Branches

To compute a branch of solutions (i.e., perform numerical continuation), we use pseudo-arclength continuation, which parameterizes a branch of solutions by their (approximated) arclength from an initial point [3], in combination with matrix–free Newton–Krylov methods [40, 41]. Pseudo-arclength continuation is an iterative process where the input at each step is: (i) the solution from the previous iteration $x_o$ and $p_o$ and (ii) the tangent direction at the previous step $\dot{x}$ and $\dot{p}$. In EF/DDDAS protocols and indeed many equation aware bifurcation studies, the tangent direction is approximated using the difference between the last two solutions on the branch. To obtain the first solution on the branch, a common strategy is to fix the value of $p$ and run the system forward in time until it converges to an equilibrium point adjust the values of $p$ slightly to find another solution, and use these solutions to compute the four needed quantities.

At each step of the process, a new solution on the branch is computed by solving the following system of equations:

$$x = F(x, p)$$
$$\dot{x}^T(x - x_0) + \dot{p}^T(p - p_0) = \Delta s \quad (2)$$

where $\Delta s$ is a user-determined step size that contains the approximated arclength each iteration covers. If one defines

$$f = \begin{bmatrix} F(x, p) - x \\ \dot{x}^T(x - x_0) + \dot{p}^T(p - p_0) - \Delta s \end{bmatrix} \quad (3)$$

Then the next solution and its corresponding parameters can be computed by supplying $f$ to a Newton-Krylov-GMRES method. The value of $F(x, p)$ is determined by running an experiment/black box simulation and recording the output state. All the other quantities are known in advance.

If a solution to this system can be found within a given tolerance, the values of $x_0$, $p_0$, $\dot{x}$, and $\dot{p}$ are updated. If a solution cannot be found, then either the step size, $\Delta s$, is too large or the noise in the experiment is simply too large to obtain accurate solutions in this part of parameter space. In either case, $\Delta s$ is decreased and the process is repeated. If $\Delta s$ becomes too small, then the process terminates.

### 2.1.2 Linear Stability Analysis

When a solution has been computed, another useful analysis is to determine the stability of the solution. This too can be accomplished using matrix-free methods and in particular the Implicitly Restarted Arnoldi Method (IRAM) [43]. The IRAM uses Jacobian-vector products, which are approximated using Eq. 1 to construct a Krylov subspace that approximates the largest eigenvalues and eigenvectors of the Jacobian. The largest eigenvalue (in magnitude) determines the spectral stability of the fixed point and least stable direction.

### 2.1.3 Estimating the Unstable Manifold

If a solution is found to be unstable, EF/DDDAS protocols can then be used to compute low-dimensional unstable manifolds. This process is relatively slow compared to the other steps so is typically not performed on every unstable solution.

There are many techniques for computing unstable manifolds [3], but in EF/DDDAS settings techniques like [44] are particularly useful because they do not require any derivatives of the system under examination. Instead, the inputs to the method are the unstable fixed point $x_0$, which is obtained from pseudo-arclength continuation, and the unstable eigenvector, $v_0$, from IRAM.

The implemented technique constructs a piecewise-linear approximation, $W_{pl}^u(x_0)$, of the true unstable manifold $W^u(x_0)$ in an iterative fashion. The initial point on the unstable manifold is $x_0$ and the first point is $x_1 = x_0 + \epsilon v_0$ where $\epsilon$ is a small perturbation. After $N$ iterations of the algorithm, $W_{pl}^u(x_0)$ is defined by $N+1$ mesh points $x_0, x_1, \ldots, x_N$ and consists of $N$ line segments between subsequent pairs of mesh points.

To extend $W_{pl}^u(x_0)$, the objective is to find a point in $\hat{x} \in W_{pl}^u(x_0)$ such that $F(\hat{x}, p)$ is a distance $\Delta_{N+1}$ away from the last point on the manifold. The value of $\Delta_{N+1}$ must be provided by the user but can be automatically reduced by the algorithm if it is too large [44]. Because this process computes one-dimensional unstable manifolds, this results in a one-dimensional search process to find the $\hat{x}$ such that $(1 - \varepsilon)\Delta_{N+1} \leq \|F(\hat{x}, p) - x_N\| \leq (1 + \varepsilon)\Delta_{N+1}$ where $\varepsilon$ is an uncertainty factor typically taken to be 0.2 and solved using the bisection algorithm. Because the bisection algorithm is derivative free, all that is required is the image of $\hat{x}$, which is computed by running a simulation/experiment with $\hat{x}$ as the initial condition. Once a valid point is identified, it is added to the unstable manifold and the process repeats.

The algorithm described above focused on one-dimensional unstable manifolds. For systems with unstable manifolds of dimension greater than one, one can supply linear combinations of unstable eigenvectors, compute 1D unstable manifold for each combination, and then later combine them to build a higher dimensional unstable manifold. Due to the computational/experimental cost this entails, in practice EF/DDDAS explorations of unstable manifolds are restricted to low-dimensional unstable manifolds.

This section has provided examples of algorithms to solve three primary tasks in bifurcation studies in an EF/DDDAS way. In order for these algorithms to be applied to black-box or experimental systems, one must use matrix-free methods that either only use function values or use function values and Jacobian-vector products. As a result, while many of the tools designed for numerical bifurcation studies (e.g., AUTO) are not applicable, the underlying techniques they implement (e.g., pseudo-arclength continuation) can be adapted to EF/DDDAS studies through relatively small changes to the solution process (e.g., Newton-Krylov methods instead of directly solving linear systems).

### 2.2 Experimental Setup

The objective of this study is to analyze the dynamics of *dark breathers* that appear in an experimental damped-driven chain of granular particles, a so-called engineered granular chain (EGC), described in [45]. A photograph of the experimental setup containing the EGC is shown in Fig. 1 where each granular particle is one of the metal beads shown in the figure, and the EGC is the entire string of beads. A *discrete breather* is a time periodic solution that is localized in space. The most common type of breathers in the literature are *bright breathers*, where the solution has a localized area of high-amplitude and decays to zero outside of that region. In this application, the focus is on *dark breathers* that have localized areas of low-amplitude. As the name implies rather than being *bright* (high-amplitude) compared to the background these solutions appear to be a region that is *dark* (low-amplitude) compared to the background.

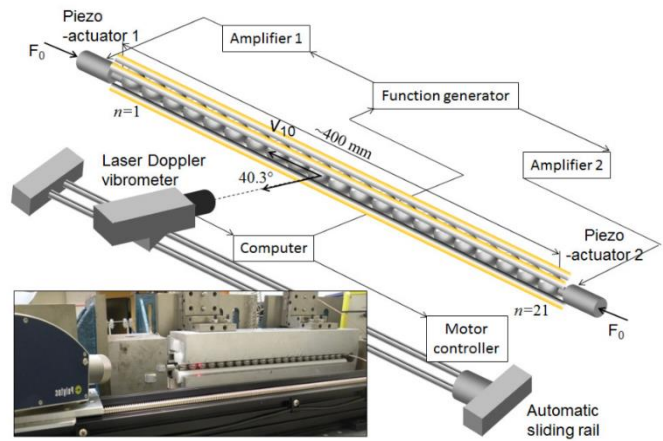

**Fig. 1.** Schematic and photograph of the experimental setup. This figure is reproduced from Ref. [45].

The experimental setup in **Fig. 1** does not allow arbitrary initial conditions to be specified. Instead, dark breathers are generated using ***destructive interference***, by initializing the beads in a quiescent state (i.e., no initial velocity) and actuating both ends of the chain in an *out-of-phase* fashion [46, 45] [17] [20]. By controlling the forcing frequency, $f_b$, one can determine the oscillation frequency of the resulting dark breather. Furthermore, because the number of beads is odd, the out-of-phase actuation will induce a vanishing oscillation amplitude at the central bead, generating the center dip often associated with dark breathers.

In the examples that follow, the system consists of a 21-bead granular chain, with piezoelectric actuators on both ends –to provide the forcing– and a laser Doppler vibrometer to non-intrusively measure the velocity of the beads that comprise the chain. Each bead is a chrome steel sphere (with radius $R = 9.53mm$, Young's modulus $E = 200GPa$, Poisson ratio $\nu = 0.3$, and mass $M = 28.2g$). Four polytetrafluoroethylene rods allow free axial vibrations of the particles while simultaneously preventing lateral motions. The entire chain is compressed with a static force of $F_0 = 10N$ as in [47].

## 3 An Equation-Aware Bifurcation Study

Before focusing on an EF/DDDAS type study**, in this section the standard approach for conducting a bifurcation study of the system in Fig. 1 is presented**. First, a set of governing equations (a model) is developed. Next, that model is calibrated and validated using experimental data. Finally, the model is utilized to perform the desired bifurcation study. This section highlights the standard procedure and presents the results that will be compared to the EF/DDDAS approach applied to the same system in the next section.

### 3.1 Governing Equations for a Damped-Driven Granular Chain

The model chosen is a standard granular crystal model [1] with a simple description of the dissipation [11] and out of phase actuators on the left and right boundaries:

$$u_0 = \alpha \cos(2\pi f_b t)$$
$$M\ddot{u}_n + \frac{M}{\tau}\dot{u}_n = A[\delta_0 + u_{n-1} - u_n]_+^{\frac{3}{2}} - A[\delta_0 + u_n - u_{n+1}]_+^{\frac{3}{2}} \quad (4)$$
$$u_{N+1} = -\alpha \cos(2\pi f_b t)$$

where $N = 21$ is the number of beads in the chain, $u_n(t)$ is the displacement of the $n^{th}$ bead from the equilibrium position at time $t$, $A = \frac{E\sqrt{2R}}{3(1-\nu^2)}$ is a constant determined by the bead composition and geometry, $M$ is the bead mass and $\delta_0$ is an equilibrium displacement induced by a static load $F_0 = A\delta_0^{3/2}$. The quantity $[x]_+ \triangleq max(0, x)$, indicates that adjacent beads exert forces *only if they are in physical contact*. $\tau$ is the time scale associated with the dissipation, $\alpha$ is the amplitude of the actuation, and $f_b$ is the frequency of the actuation. The numerical value of most of these parameters can be obtained from the physical properties of the beads, but the value for the damping ($\tau = 5$ ms) was empirically determined from experimental observations [45]. In both the mathematical model and the experimental setup, $\alpha$ and $f_b$ are free to vary.

The dark breathers of interest appear to have large basins of attraction (i.e., the same dark breather state can be obtained from many initial conditions), making them observable both experimentally and numerically when an appropriate actuation frequency is chosen. Starting from a quiescent (zero) initial condition, the propagation of plane waves in the linear regime and their subsequent destructive interference leads to the spontaneous emergence of the dark breather (i.e., the quiescent state lies within the basin of attraction of a dark breather). In order to ensure the robust *experimental* observation of a dark breather, and avoid the onset of transient, large amplitude traveling waves [1], the actuation amplitude is tuned to increase linearly from zero to the desired amplitude α over eight forcing periods.

### 3.2 Model Validation

With the model and its parameters in hand, the next task is validation, by comparing experimentally obtained dark breathers with numerically computed ones using identical sets of parameters (see Fig. 2). In this example, there is good agreement between the experimentally observed and simulated dynamics. As a result, one could reasonably analyze this system using model-based techniques. Ultimately, the need for EF/DDDAS protocols is driven by systems where good agreement between models and experiments cannot be obtained so explicit equations for the dynamics are not known.

In this example, however, having explicit equations enable us to ensure the EF/DDDAS solutions agree with the solutions obtained given an explicit model.

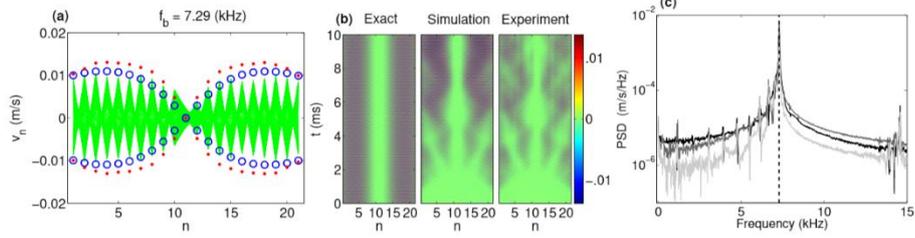

**Fig. 2**: (a) Experimentally measured velocities versus bead number for $f_b = 7.29\ kHz$ and $a = 0.2\ \mu m$. The entire time series of each bead location is shown as a superposition (green lines). Extrema as predicted by the simulation for the first $10\ ms$ (red points) and from the numerically exact dark breather (blue circles) are also shown. (b) Space-time contour plots of the exact breather, a transient starting from zero initial data, and the experimental evolution, also from zero initial data, leading to the same state. Color intensity corresponds to velocity $(m/s)$. (c) Experimentally measured power spectral density $(m/s/Hz)$ of bead 1 (darkest line), bead 5 (dark line) and bead 10 (lightest line) for an actuation amplitude of $a = 0.2\ \mu m$. The vertical dashed line corresponds to a driving frequency of $f_b = 7.29\ kHz$. This plot is reproduced from [45].

To compute periodic orbits in this damped-driven system, the computation is performed for fixed points of the numerically realized *stroboscopic map* $\boldsymbol{F} = \boldsymbol{u}(T_b) - \boldsymbol{u}(0)$, where $\boldsymbol{u}(0)$ is the initial state vector and $\boldsymbol{u}(T_b)$ is the solution of the equations of motion at time $T_b$. The Jacobian of $\boldsymbol{F}$, which is used in the Newton iterations, is of the form $\boldsymbol{V}(T_b) - \boldsymbol{I}$, where $\boldsymbol{I}$ is the identity matrix, $\boldsymbol{V}$ is the solution to the variational equation $\dot{\boldsymbol{V}} = \boldsymbol{DF} \cdot \boldsymbol{V}$ with initial condition $\boldsymbol{V}(0) = \boldsymbol{I}$ and $\boldsymbol{DF}$ is the Jacobian of the equations of motion evaluated at $\boldsymbol{u}$. To be noted that the breather frequency and the actuation frequency are both $f_b = 1/T_b$ by construction.

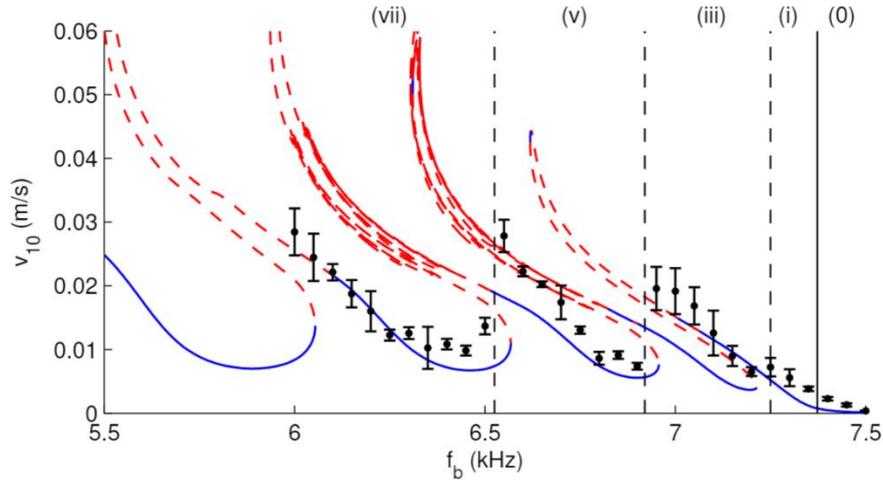

**Fig. 3:** Bifurcation diagram illustrating the various dark breather/multibreather branches. Specifically, the maximum velocity of bead 10 versus frequency $f_b$ is shown with $a = 0.2\ \mu m$.

The smooth curves correspond to the numerically exact dark breathers (solid blue lines correspond to linearly stable solutions and red-dashed lines are unstable regions). The markers represent the experimentally measured mean values with standard deviations given by the error bars, obtained using four experimental runs.

Fig. 2 (a-b) shows a numerical and experimental dark breather excitation for values of $f_b = 7.29\ kHz$ and $a = 0.2\ \mu m$ (which were chosen to match the experimental forcing frequency and amplitude) and the corresponding numerically obtained solution at those parameter values. The strong structural similarity in space to an exact dark breather (see, e.g., panel (a)) suggests that both are near the same stationary state after $10 ms$. Furthermore, it also implies that the basin of attraction of the dark breather state does contain the quiescent solution in both the experiment and the mathematical model. To determine whether the experimental data is time periodic, the power spectral density (PSD) plot shown in panel (c) is considered. The strong spike in the PSD near $f_b = 7.29\ kHz$ implies the experimental system is also converging towards a periodic orbit with this frequency. The detailed comparison of the space-time evolution of the numerically exact solution, simulation, and experiment in panel (b) shows that after the initial transient stage of the dynamics, a dark breather is indeed formed. Notably, the maximum strain, $|\boldsymbol{u_n} - \boldsymbol{u_{n+1}}|$, of the solution is about 60% of the precompression, confirming that the structures reported here are a result of the nonlinearity of the system.

To validate the model in the multibreather (i.e., $n$-dip solution) regime, the full bifurcation diagram of the dark breathers shown in Fig. 3 and the corresponding representative profiles/evolutions illustrated in Fig. 4, is examined in order to validate the model in the multi-breather (i.e., $n$-dip solution) regime. The fold on each lobe represents the collision of two breather families (i.e., the $n$- and $(n+2)$-dip solutions). Due to the structure of the branch, the transition from one family of dark breathers to the next (as, say, $f_b$ is reduced) is not smooth: that is saddle-node bifurcations (i.e. collision and disappearance of two equilibria in dynamical systems) cause the solution to "jump" from family to family. The saddle-node can be seen in Fig. 3, where the regions labeled (vii), (v), (iii) and (i) correspond, respectively, to the number of density dips seen in the (experimental) space-time contour plot. The label (0) corresponds to the surface discrete (bright) breathers localized near the boundaries rather than at the center of the domain. The experimentally measured solutions are indicated by black markers with error bars in Fig. 3. The model (Eq. 4) used here for the dark breathers thus derived, yields, at least qualitatively, the correct type of multibreather at the appropriate forcing frequency.

On a more detailed level, the transient development of a given multi-breather for a fixed forcing frequency is examined. For example, in region (vii) at $f_b = 6.35\ kHz$ a solution with 7 dips emerges (see Fig. 4a). Notably the visible agreement between the *exact* solution (obtained by numerically locating the fixed point via Newton's method), the *simulation* (obtained by numerically integrating (4) starting from the quiescent solution), and the experimental results taken over the same time window. Both the simulation and experiment approach the same solution with similar transients starting with quiescent initial conditions. Similar agreement is observed with 5 dips (e.g., for $f_b = 6.80\ kHz$ in Fig. 4b), 3 dips (e.g., for $f_b = 7.15\ kHz$ in Fig. 4c), and ultimately the 1 dip solution already mentioned in Fig. 2.

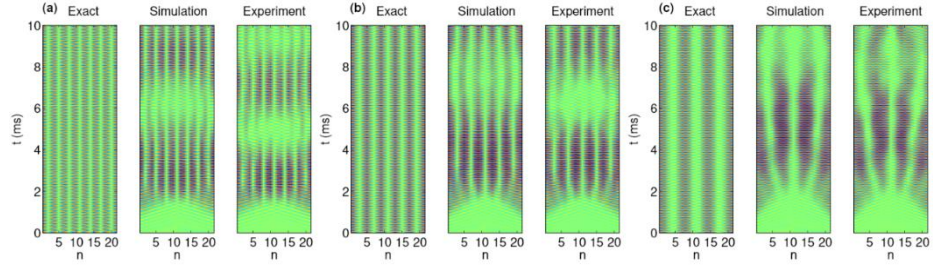

**Fig. 4:** Space-time contour plots of the numerically obtained dark breather, as well as a transient simulation and the experimental observation of the dynamic evolution leading to it, with a=0.2 μm. (a) a 7-dip solution with $f_b$ =6.35 kHz. (b) a 5-dip solution with $f_b$ =6.8 kHz. (c) a 3-dip solution with $f_b$ =7.15 kHz. Color intensity corresponds to velocity (m/s); the color legend of each panel is the same as in Fig. 2. This plot is reproduced from [45].

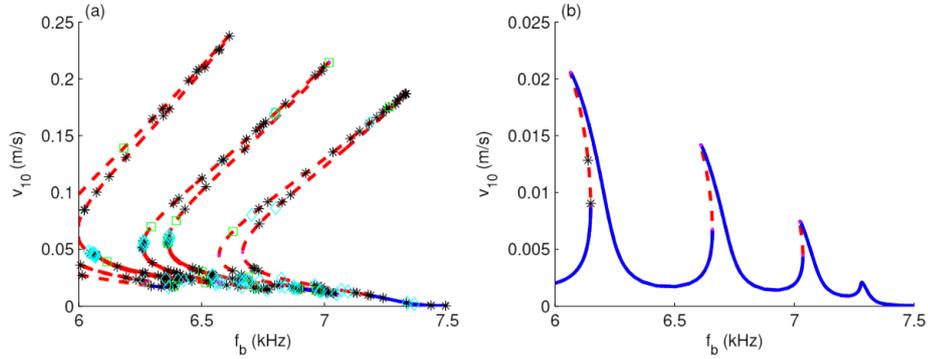

**Fig. 5:** (a) Bifurcation diagram computed at a larger forcing amplitude of $a = 0.5$ μm, and (b) at a smaller forcing amplitude of of $a = 0.05$ μm. Stable branches, as defined in [10], are denoted by solid (blue) lines and unstable segments of the branch are denoted by dashed (red) lines. Saddle-node bifurcations are denoted by (magenta) dots, Neimark–Sacker bifurcations by (black) stars, period-doubling bifurcations by (green) squares, and pitchfork bifurcations by (cyan) diamonds. Note that at both forcing amplitudes the same qualitative structure persists, but the frequency at which a given n-dip solution occurs is shifted.

Lastly, it was noted that the system modeling could also "dial in" dark breathers by changing the forcing amplitude. To determine whether that occurs in the model, we performed two additional bifurcation studies in Fig. 5, showing bifurcation diagrams obtained at different forcing amplitudes ($a = 0.5$ and $0.05\,\mu m$) respectively. The same lobe-like structure in frequency appears; however, the frequency range required to generate a given multibreather will shift, and as a result, is possible, within reason, to dial in a given multibreather by changing the forcing amplitude as well.

In summary, in this section, a standard bifurcation study of a model of a physical engineered granular chain (EGC) was performed. First, the model was defined, followed by verification; next, the model was validated using the available data; and finally, the resulting model was used in the parameter regimes where it was validated.

The EGC system is very rich from a dynamical systems perspective; it clearly demonstrates snaking, and has many secondary, tertiary, and higher branches. Furthermore, the bifurcation study uncovers information useful and provides a tool for validating the underlying model by allowing matching where qualitatively different types of behaviors occur. Clearly, if one could use EF/DDDAS protocols to "skip" the entire step of model creation and validation, this entire step could be circumvented.

## 4 An EF/DDDAS Study of the Dynamics

Using the EF/DDDAS computational tools described in Section 2.1, the bifurcation diagram in Fig. 6 is produced; it is the EF/DDDAS analog of the bifurcation diagram in Fig. 3. First, is noted the quantitative agreement between the figures; that is, the collection of matrix–free tools is able to reproduce the branch and bifurcations that standard tools could, despite the fact the accuracy of the EF/DDDAS methods are typically lower than the tolerances used in tools like AUTO. Although an individual solution may be computed to lower accuracy, the change is solutions over the course of the branch is much greater than the error in any individual solution. Therefore, the shape of the branch is recovered even if the error at every step of the process is higher. In this example, the structure a complex "snaking" [48, 49, 50, 51, 52, 53] structure that has received considerable recent attention in settings such as nematic liquid crystals [54] and classical fluid problems such as Couette flow [55].

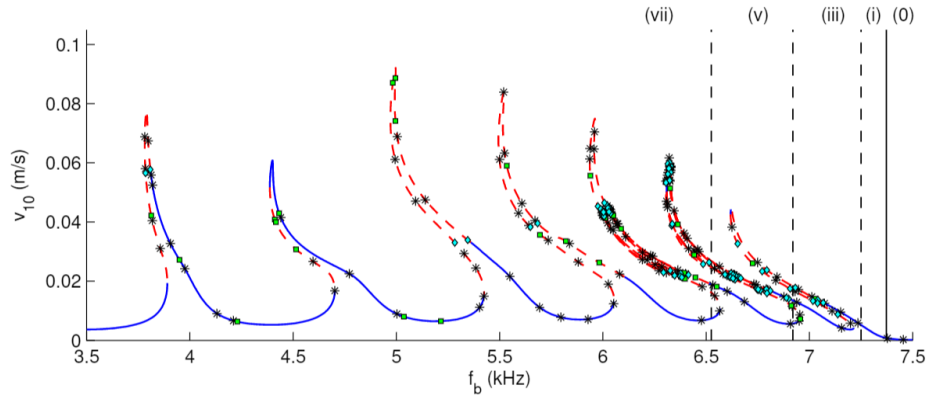

**Fig. 6:** The branch of solutions computed using the EF/DDDAS approach, which should be compared with Fig. 3. The solid (blue) regions indicate that the real Floquet multipliers are within (or on) the unit circle. A dashed (red) line is shown otherwise. The symbols indicate the presence of a bifurcation: black stars are Neimark–Sacker bifurcations to $T^2$ tori, green squares are period doubling bifurcations, and cyan diamonds are pitchfork bifurcations. Here, we have only plotted half the total number of torus bifurcations to avoid obscuring the plot. The dashed lines indicate the regions where the n-dip solutions appear as shown in Fig. 3.

Next, the bifurcation study presented above is extended to describe secondary branches of solutions (i.e., those generated by the bifurcations that occur on the main branch). To determine when a bifurcation has occurred, the bordered matrix techniques used by AUTO [3] are adopted, by solving the necessary linear system of equations in a matrix–free fashion. In this way, the EF/DDDAS methods can locate pitchfork and period doubling bifurcations, identify Neimark–Sacker bifurcation [56] points, and directly track the *slow* complex-conjugate pairs of eigenvalues to determine when they cross the unit circle. The former two are the points on the main branch where the secondary branches of periodic solutions are born, whereas the Neimark–Sacker points suggest the existence of quasi-periodic solutions. The next section presents a brief example of the secondary branches generated by each of these bifurcations.

### 4.1 Pitchfork Bifurcations

Many of the secondary branches arise at pitchfork bifurcations where a symmetry that was contained in the original branch of solutions is broken and two additional branches of solutions are generated. Pitchfork bifurcations can also appear as part of what we call *a pitchfork loop*, see Fig. 7. Pitchfork loops contain two branches of solutions that are generated via a pitchfork bifurcation on a branch of solutions and are later destroyed via another pitchfork bifurcation on the same branch. Other bifurcations may occur on these branches, and the prototypical example of a pitchfork loop consists of a pair of subcritical pitchfork bifurcations and two pairs of saddle-node bifurcations on a single, secondary branch of solutions. As a result, pitchfork loops physically appear as a small region of parameter space where a symmetry in the underlying solution branch vanishes only to later reappear.

Ultimately, the hallmark of a pitchfork bifurcation or pitchfork loop is the loss of a symmetry that previously existed in the system. Due to the out-of-phase forcing at the ends of the EGC, most solutions have velocities that are odd functions of the bead number (i.e., $\dot{u}_1(t) = -\dot{u}_N(t)$, $\dot{u}_2(t) = -\dot{u}_{N-1}(t)$, and so forth) with the center bead remaining stationary. However, solutions on the pitchfork loops shown in Fig. 7 lose this symmetry. To demonstrate this, note that the vector of bead velocities can always be decomposed into an even and odd components:

$$\begin{bmatrix} \dot{u}_1(t) \\ \dot{u}_2(t) \\ \vdots \\ \dot{u}_{N-1}(t) \\ \dot{u}_N(t) \end{bmatrix} = \dot{\boldsymbol{\alpha}}(t) + \dot{\boldsymbol{\beta}}(t) = \begin{bmatrix} \dot{\alpha}_1(t) \\ \dot{\alpha}_2(t) \\ \vdots \\ -\dot{\alpha}_2(t) \\ -\dot{\alpha}_1(t) \end{bmatrix} + \begin{bmatrix} \dot{\beta}_1(t) \\ \dot{\beta}_2(t) \\ \vdots \\ \dot{\beta}_2(t) \\ \dot{\beta}_1(t) \end{bmatrix},$$

where $\dot{\boldsymbol{\alpha}}$ is the component of velocity that is an odd function of bead number and $\dot{\boldsymbol{\beta}}$ is the component of velocity that is an even function of bead number. Fig. 7a shows the solutions branches in terms of the velocity of the $10^{th}$ bead while Fig. 7b shows the same solution branches but the *y*-value is given by $v^T \beta$ where $v$ is a random vector chosen such that $v^T \alpha = 0$. As shown in Fig. 7b, the original branch of solutions contain no velocity components that are even functions of the bead number. However, solutions in the pitchfork loops break this symmetry and introduce velocity components that are even functions of the bead number.

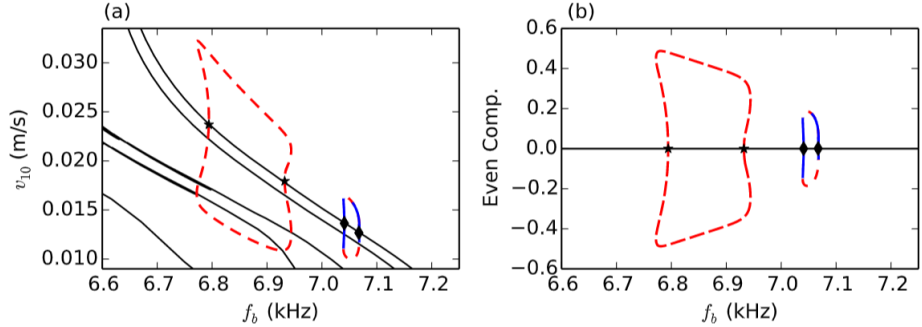

**Fig. 7:** A prototypical example of a pitchfork loop. The plot on the left (a) shows the branch using the same projection as in Fig. 6, while the plot on the right (b) illustrates the symmetry breaking that occurs by projecting the velocity of the beads onto a random even vector (i.e., a vector, $v$, where $v_1 = v_{21}$ and so forth). The (black) markers denote the pitchfork bifurcations of the loop. In particular, the main branch has effectively no even components while the upper and lower branches in the pitchfork loop do.

Experimentally detecting pitchfork loops from raw sensor measurements (i.e., by observing a timeseries of sensor measurements) can be challenging. This is due in large part by the fact that solutions on a pitchfork loop often remain close to solutions on the (now unstable) main branch in terms of standard metrics like the $L_2$-norm. As a result, both equation-aware and EF/DDDAS protocols that otherwise perform well may miss pitchfork loops simply because they have a smaller impact than the coiling (i.e., the sequence of saddle-node bifurcations in Fig. 3) that occurs on the main branch. However, they can be detected due to the symmetry breaking associated with them. In this example, they were detected due to the even in bead number component of velocity that appears in both the solution and the largest eigenvalue in magnitude similar to how they are detected in equation aware packages such as AUTO [3].

### 4.2 Period Doubling Bifurcations

The second set of branches is generated by repeatedly occurring period doubling bifurcations. Many of these secondary branches are initiated in regions where the main branch is already unstable, usually due to an earlier Neimark–Sacker bifurcation. These solution branches are complex and exist for large intervals of $f_b$, but all of the solutions identified were unstable due to a pre-existing instability. Ultimately, if one were able to initialize the system at will, it would be possible to observe these solutions for short periods at a number of different forcing frequencies, but because they are unstable they will not appear when starting from a quiescent solution.

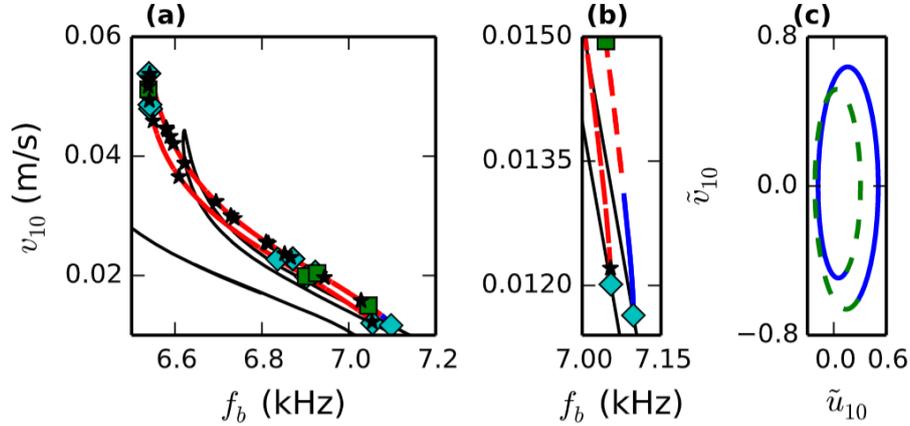

**Fig. 8:** (a) An example of a period-two branch of solutions is shown in blue and red; the solid (blue) regions (bottom right – visible when zoomed in panel (b)) indicate that the real Floquet multipliers are within (or on) the unit circle so that small perturbations to the solution will not grow in time [56]. A red line is shown otherwise. The thin (black) curve denotes the period-one branch of solutions, and the markers indicate pitchfork bifurcation (cyan diamonds), period-doubling bifurcations (green squares), and Neimark–Sacker bifurcations (black stars). (b) A zoom of panel (a) showing how each end of the period-two branch connects to the main branch. (c) A phase plane plot of the (nondimensionalized) velocity versus position of a sample period-two solution away from the bifurcation point. The solid (blue) line is the first half of oscillation (one forcing period), and the dashed (green) line is the second half of the oscillation.

A prototypical example of a secondary, period-two branch is shown in Fig. 8. This period two branch acts as a "bridge" between two sides of a single lobe of period-one solutions as indicated by the pair of connections at the start and end of the period-two branch of solutions. The period-doubling bifurcations on the main branch do create two secondary branches of solutions, but the ordinate in Fig. 8, which is the maximum velocity of the tenth bead, causes the two branches to appear practically identical.

As in the main branch, the period-two branch contains its own set of bifurcations including additional branch points, further period doubling bifurcations, and Neimark–Sacker bifurcations, each of which could create tertiary period-two, period-four, or quasi-periodic solutions respectively. However, none of the tertiary period-two or period-four branches we examined were born stable at their associated bifurcation points, nor was stability regained in the tertiary branch. Because none of these higher period solutions are stable, they will not appear in experiments unless very specific initial conditions are chosen. As a result, while these solutions exist they are not expected to play a significant role in the dynamics of dark breathers.

### 4.3 Quasi-periodic Solutions

Due to the presence of many Neimark–Sacker points, it is natural to expect quasiperiodic solutions to exist in this system. A quasi-periodic solution is a continuous time trajectory on a torus in phase space (in the stroboscopic map, these trajectories appear as invariant circles). Like the periodic orbits born out of supercritical Hopf bifurcations, supercritical Neimark–Sacker bifurcations generate stable tori in the neighborhood of the bifurcation point [56]. Algorithms for approximating invariant circles exist (and can even be computed in a - *matrix-free* fashion) [57], but computing branches of quasi-periodic solutions is outside the scope of this chapter. Instead, in the neighborhood of the Neimark–Sacker bifurcation, the invariant circle is approximated by recording the stroboscopic iterates many times as demonstrated in Fig. 9. Fig. 9 gives two examples of such a computation using $10^5$ successive iterations of the stroboscopic map for each value of $f_b$. The figure itself shows the next $10^4$ iterates of the map with one marker every 10 iterates. Note that this number of iterations was chosen for the sake of visualization and could be adjusted with no change to the results.

At $f_b = 7.487\ kHz$, we have identified what appears to be a quasi-periodic orbit as shown in Fig. 9 (a-b). Panel (a) shows the displacement and velocity of the $10^{th}$ bead at the start of every forcing period. The blue dots denote the state of the system (in this projection) after every 10 iterates of the stroboscopic map, while the green dots denote the last 20 iterates of the map; and lastly, the black star is the period-one solution at this frequency. The invariant circle itself appears to be stable even in the face of large perturbations.

Fig. 9b shows the relative distance between two nearby trajectories as a function of iterations of the stroboscopic map. Trajectories that start not too far apart will converge slightly as the perturbed solution is attracted back to the invariant circle and will remain close for all future times.

The behavior at $f_b$ = 7.487 kHz should be contrasted with what occurs at $f_b = 7.462\ kHz$, which is shown in Fig. 9 (c-d). Although the dynamics shown in panel (c) might be thought to lie on a higher order torus, say a $T3$–torus, the plot of the distance between nearby initial conditions in panel (d) indicates that it is more likely on a chaotic attractor as nearby trajectories diverge after sufficiently long periods of time. This can be explained by the Ruelle-Takens-Newhouse route to chaos, which occurs when a quasi-periodic solution (a solution on a $T^2$-torus) undergoes another Hopf bifurcation (yielding a solution on a $T^3$-torus), and is based on the observation that a constant vector field on a $T^3$- torus can be perturbed by an arbitrarily small amount to produce a chaotic attractor [58].

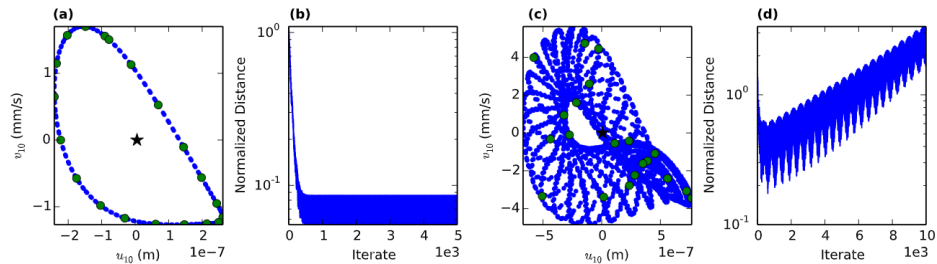

**Fig. 9:** (a) A representative invariant circle created by plotting several thousand iterates of the map F with $f_b = 7.487\ kHz$; the small (blue) dots indicate the velocity of the 10th bead recorded every $10/f_b$ seconds, the large (green) dots denote the last twenty iterates of the map, and the star (black) denotes the unstable period-one orbit. (b) The log of the normalized distance between two nearby initial conditions as a function of time; though the exact distance is time dependent, trajectories on the invariant circle that start near each other remain near each other. (c) A demonstration of the stroboscopic map with $f_b = 7.462\ kHz$ colored like the leftmost plot; if more iterations are plotted, the trajectory will eventually "fill in" the region. (d) Pairs of nearby trajectories at $f_b = 7.462\,kHz$ will separate after a sufficiently large number of iterations, which suggests the presence of a chaotic attractor. The figure is reproduced from [45].

### 4.4 Computing Unstable Manifolds

The EGC example discussed here is an example of a forced dynamical system due to the out-of-phase actuation at the ends of the chain. For some combinations of forcing frequency and amplitude, the EGC will be *phase-locked* with the forcing (i.e., the frequency of the response will be a rational multiple of the frequency of the forcing). This results in periodic solutions when the system is viewed using a stroboscopic map. In other regions, however, the ratio between system frequency and forcing frequency is irrational, which results in quasi-periodic solutions. The regions where phase-locking occurs are called Arnold tongues [59].

The objective of this section is to analyze the transitions between phase-locked and quasi-periodic behavior at large forcing amplitudes in an EF/DDDAS manner based on the unstable manifolds associated with phase-locked solutions. As mentioned in Section 2.1, computing high-dimensional unstable manifolds is a challenge for the current tools. Fortunately, despite the complicated dynamics inherent in the EGC, there are many regimes where only one or two Floquet multipliers have exited the unit circle. As a result, though the solution itself may be 42-dimensional, the unstable manifold is only 1- or 2-dimensional, which is small enough to make computations of the unstable manifold (and its visualization) feasible.

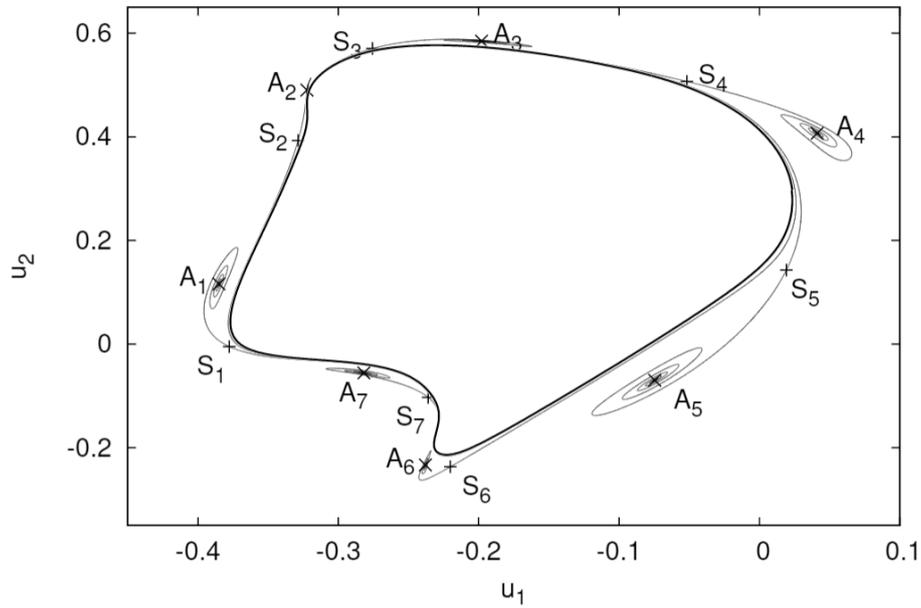

**Fig. 10:** Plot of the one-dimensional (Poincaré map) unstable manifold of a saddle dark breather with $a = 0.278\,\mu m$ and $f_b = 7.14\,kHz$ expressed in terms of the (nondimensionalized) displacement of beads 1 and 2. The 7 saddle points (belonging to the same period–7 orbit) are labeled $S_1$–$S_7$; in this image, there are two sets of attractors, the stable spirals labeled $A_1$–$A_7$

(belonging to another period–7 orbit) and the stable invariant circle indicated by the thick line. Close to the tip of the Arnold tongue, the $A_i$ are nodes, but at this point, the original torus has broken, and a new invariant circle has appeared.

In particular, a typical solution within an Arnold tongue consists of $N$ saddle points and $N$ nodes when viewed in terms of the stroboscopic map [60]. The value $N$ depends on the relationship between the forcing and response frequencies. In Fig. 10, for example, the ratio of the system frequency to the forcing frequency in this example is $\frac{1}{7}$, so the phase-locked solutions are period 7 solutions in the stroboscopic map and $N = 7$. The saddles and nodes are linked via the unstable manifolds of the saddles in a *heteroclinic connection*.

Given a fixed forcing amplitude and changing forcing frequency, phase-locked solutions are destroyed when the saddles and nodes combine in a saddle-node bifurcation [61]. However, Fig. 10 considers the case where the frequency is fixed by the forcing amplitude is too large. In this regime, the loss of phase-locking is more complex. For example, Fig. 10 still contains a set of seven saddle points ($S_1$-$S_7$) whose unstable manifolds, shown in light gray, connect with a set of seven fixed points. However, rather than nodes these points are stable spirals ($A_1$-$A_7$) as shown in the figure by the spiral pattern of the unstable manifold makes as it approaches these points. Furthermore, the unstable manifold of the saddles also connects with a quasi-peiodic solution indicated by the black curve. The collision of equilibrium points such as the saddles and spirals with a larger invariant such as the quasi-periodic solution is indicative of a global bifurcation such as a heteroclinic bifurcation.

Physically, the result of a global bifurcation is that the transition between phase-locked and non-phase-locked is not as clean is it would be in the fixed-amplitude varying frequency case. In that case, local bifurcations between the saddles and nodes would eliminate the phase-locked, period-7 solution. In this case, however, that solution still exists ($A_1 - A_7$) but is near a quasi-periodic solution, and indeed, the unstable manifold of the saddles can lead to either type of solution. As a result, given the same forcing frequency and amplitude, it is possible to obtain either solution type, and because they are close in state-space, perturbations could cause one solution type to transition to the other.

Bifurcation studies are an important step in understanding dynamical system. Given access to explicit governing equations, packages like AUTO have automated many of the necessary computational tasks. In this section, it has been demonstrated the EF/DDDAS protocols can perform three key tasks in a bifurcation study: (i) computing a solution branch, (ii) assessing the linear stability of solutions, and (iii) computing unstable manifolds. Ultimately, using a black-box simulator, EF/DDDAS protocols were successful in extracting the same information obtained from an equation-aware study using AUTO on a non-trivial system that includes a snaking branch solutions and multiple non-trivial bifurcations despite having less knowledge about the underlying system.

## 5 Conclusion

Performing bifurcation studies on models of a physical system is an important part of obtaining both an intuitive and a quantitative understanding of that system's dynamics. This chapter highlights two approaches for performing such a study. First, one could formulate a model, use experimental data to validate that model, and then use the resulting model in combination with software such as AUTO [3] to compute the bifurcation diagram. Alternatively, one could use EF/DDDAS protocols to compute the bifurcation diagram by running judiciously initialized experiments (or in this case, black-box *numerical simulation*s) to provide the information needed to locate fixed points and determine their stability on the fly. The algorithms underpinning this are the so-called *matrix-free* algorithms such as Newton-GMRES (for finding fixed points) and the Implicitly Restarted Arnoldi method (IRAM) for computing eigenvalues and eigenvectors. It should be noted that EF/DDDAS protocols make use of the same mathematical structure that the standard methods accomplish. Indeed, pseudo-arclength continuation and the bordered matrices used to locate fixed points can all be directly imported in the EF/DDDAS framework, which allows computing the bifurcation diagrams equivalent to (and quantitatively agreeing with) those computed using AUTO.

The EF/DDDAS bifurcation study presented could be performed without knowledge of the governing equations of the engineered granular chain. The DDDAS concept does not necessitate Jacobian estimation but only input/output information and allows reliable construction of the bifurcation diagram.

One of the main limitations on the current approach is that it is assumes initial conditions can be specified at will. This is advantageous because it enables EF/DDDAS protocols to leverage the techniques and analysis of matrix-free linear algebra. However, it is also limiting. For the example of the EGC, it is possible to experimentally prescribe several features of the initial condition (e.g., the state of the beads at the boundaries), but prescribing initial velocities/displacements in the interior of the chain is difficult. Similarly, unstable parts of solution branches contain useful information, and may in fact become stable at other parameter regimes, yet without the ability to specify initial conditions these solutions must be stabilized through some form of external control before they can be tracked.

A possible approach to this problem is to leverage ideas from feedback control and system identification, which are designed for cases with more limited control authority (see, for example, [62, 63, 64]). For example, in order to determine stability, methods such as Dynamic Mode Decomposition [65, 66], or the Eigensystem Realization Algorithm (ERA) [67] can be used to generate linear approximations of the underlying dynamics so as to identify unstable directions/bifurcations. If effective algorithms that require less control authority can be identified, the set of nonlinear systems to which EF/DDDAS can be applied would be greatly increased, and this area is a promising direction of future work.

## Acknowledgements

The authors would like to thank G. Theocharis for useful discussions. Support from US-AFOSR (FA9550-12-1-0332) and US-NSF (CMMI 844540, 1200319, 1310173, and1000337) is greatly appreciated. MOW acknowledges support from US-NSF DMS 1204783.## References

[1]    V. F. Nesterenko, Dynamics of heterogeneous materials, Springer, 2001.